\documentclass[12pt, aps,amsmath,showpacs,superscriptaddress]{revtex4-1}
\usepackage{graphicx,dcolumn,subfig,caption,bbm,amssymb,amsmath,ulem,indentfirst,amsthm,color,cleveref}
\usepackage{natbib}
\begin{document}
\title{Thermal chemical reactivity in Frenkel exciton–polariton cavities}


\author{Bingyu Cui}
\email{bycui@cuhk.edu.cn}
\affiliation{School of Science and Engineering, The Chinese University of Hong Kong, Shenzhen, Guangdong, 518172, P. R. China}

\author{Abraham Nitzan}
\email{anitzan@sas.upenn.edu}
\affiliation{Department of Chemistry, University of Pennsylvania, Philadelphia, Pennsylvania 19104,
USA}
\affiliation{School of Chemistry, Tel Aviv University, Tel Aviv 69978, Israel}

\date{\today}

\begin{abstract}
\noindent Hybrid light–matter states formed under strong coupling between molecular excitations and confined electromagnetic modes provide a potential route to modify chemical properties. Here we compute and compare a thermally averaged measure of molecular chemical activity for an equilibrium ensemble of molecules inside and outside a planar microcavity, explicitly accounting for the spatial distribution (and hence the in-plane wavevector dispersion) of the coupled light–matter states. Within a generalized Tavis–Cummings description, we find that the cavity-induced change in thermal chemical activity is most pronounced for small molecular ensembles (low areal density within a given cavity mode volume) and increases with the collective coupling strength (Rabi splitting), particularly at low temperatures. These results highlight the importance of the polariton dispersion and molecular-mode counting in assessing cavity modifications of thermally driven molecular reactivity.
\end{abstract}
\maketitle

\section{Introduction}
Photochemistry---the study and application of light-induced chemical processes---is central to a broad range of chemical and materials problems \cite{Kohler1995,Crim1999}. A growing body of work has examined how structured electromagnetic environments, including plasmonic nanostructures and optical cavities, can modify optical response and photochemical outcomes by reshaping the local field and the radiative density of states \cite{Dunkelberger2022}. In the weak-coupling regime, such effects underlie well-established phenomena in surface-enhanced spectroscopy and photochemistry \cite{Keresztury2001,Siebert2TO,Nitzan1981,UENO201331,Sukharev_2017}.

In the strong-coupling regime, collective coupling of many molecules to confined electromagnetic modes yields hybrid light--matter eigenstates (polaritons). While the optical signatures of strong coupling are well understood, its implications for chemical change remain under active debate. A central conceptual challenge is the mismatch between the collective character of polaritonic excitations and the local nature of molecular structural rearrangement. For example, barrier crossing along a reaction coordinate is typically a single-molecule event rather than a concerted motion of the entire ensemble. We have recently shown that this interplay can lead to an effective modification of the local potential associated with a reaction coordinate \cite{Cui2022}.

Reports of cavity-modified thermal reactivity, including effects attributed to dark-state manifolds, pose additional theoretical challenges \cite{Nagarajan2021,Hutchison2012,Dunkelberger2022}. Standard transition-state treatments based on single-mode Tavis--Cummings models generally do not reproduce such observations \cite{Thomas2016,Thomas2019,Thomas2019b,Lather2019}, and dynamical corrections have not yet yielded a unified picture \cite{Nitzan2006}, particularly regarding the reported correlation with collective coupling strength (Rabi splitting) \cite{Li2020,Feist2020}.

From a statistical-mechanical perspective, thermally activated dynamics are governed by free-energy landscapes, and understanding how strong coupling modifies the relevant free energies is therefore an essential prerequisite for assessing cavity effects on thermal reactivity. Motivated by this viewpoint and by earlier arguments comparing the free energies of bright/polaritonic and dark manifolds \cite{Scholes2020}, we extend the analysis to include the full in-plane dispersion of a planar cavity mode \footnote{Our calculations apply to localized Frenkel-type excitations, appropriate for organic materials in the strong-coupling regime.}. We then evaluate a population-based proxy for chemical activity that assumes that molecules become chemically active only upon electronic excitation, and we examine how the result depends on molecular density (mode counting) and collective coupling strength.

In what follows, energies are reported in units of $E_{xg}$ (the molecular electronic transition energy), and lengths in units of $\hbar c/E_{xg}$, with $c$ the speed of light (in the cavity medium). Unless otherwise stated, we set $\hbar = k_B = 1$.

The structure of the article is as follows: In Sec. II, we briefly recap the model of cooperative coupling in the planar cavity for molecular excitations classified as Frenkel excitons and elaborate on the associated free energy landscape. In Sec. III, we calculate and compare the thermal chemical reactivity of an ensemble of molecular clusters with or without the cavity. Finally, we summarize and conclude in Sec. IV.


\section{Statistical mechanics of exciton-polariton systems in a planar cavity}
To describe molecules laterally distributed in a planar microcavity and collectively coupled to confined electromagnetic modes, we adopt a generalized Tavis--Cummings (GTC) model \cite{Keeling2020}:

\begin{equation}
    \hat{H}_{GTC}=\sum_{\mathbf{k}}\hbar\omega_{\mathbf{k}}\hat{a}^\dagger_{\mathbf{k}}\hat{a}_{\mathbf{k}}+E_{xg}\sum_j\hat{\sigma}_j^+\hat{\sigma}_j^-+\sum_{j,\mathbf{k}}g(\hat{\sigma}_j^+\hat{a}_{\mathbf{k}}e^{i\mathbf{k}\cdot\mathbf{R}_j}+\hat{\sigma}_j^-\hat{a}_{\mathbf{k}}^\dagger e^{-i\mathbf{k}\cdot\mathbf{R}_j}),
    \label{eq:HGTC}
\end{equation}
where the operator $\hat{a}_{\mathbf{k}}^\dagger$ ($\hat{a}_{\mathbf{k}}$) creates (annihilates) a cavity-mode photon of energy $\hbar\omega_{\mathbf{k}}$ with the in-plane wavevector $\mathbf{k}_{||}=\mathbf{k}=(k_x,k_y)$, while $\hat{\sigma}_j=|g_j\rangle\langle e_j|$ and $\hat{\sigma}^+_j=|e_j\rangle\langle g_j|$ describe transitions between the lower $|g_j\rangle$ and upper $|e_j\rangle$ electronic states of molecule $j$, respectively. All molecules are characterized by the same energy spacing $E_{xg}$, coupling to the cavity field and position $\mathbf{R}_j=(x_j,y_j)$. The cavity photon dispersion is $\omega_{\mathbf{k}}=c\sqrt{\mathbf{k}^2+(2\pi n/L_c)^2}/n_c$ where the integer $n$ enumerates the cavity mode, with $L_c$ the effective cavity length (distance between mirrors), $c$ the speed of light in vacuum and $n_c$ the refractive index inside the cavity. Below we take $c/n_c=1$. It is convenient to set periodic boundary conditions for the lateral cavity geometry so that $k_x,k_y$ take discrete values $2\pi j/L,j=0,\pm1,\pm2,...$ with $L$ the lateral length. It is also convenient to take the molecules to be placed on a square lattice with lattice spacing $d=L/\sqrt{N}$ so that the molecular density per unit lateral area is $1/d^2$. Taking Fourier transform $\hat{\sigma}_j^-=\sum_{\mathbf{k}'}\hat{\sigma}_{\mathbf{k}'}^-e^{i\mathbf{k}'\cdot\mathbf{R}_j}/\sqrt{N},\hat{\sigma}_j^+=\sum_{\mathbf{k}'}\hat{\sigma}_{\mathbf{k}'}^+e^{-i\mathbf{k}'\cdot\mathbf{R}_j}/\sqrt{N}$ leads to (see details in Sec. I in the Supplementary Information (SI)) 
\begin{equation}
   \hat{H}_{GTC}=\sum_{\mathbf{k}}\hbar\omega_{\mathbf{k}}\hat{a}^\dagger_{\mathbf{k}}\hat{a}_{\mathbf{k}}+E_{xg}\sum_{\mathbf{k}'}\hat{\sigma}^+_{\mathbf{k}'}\hat{\sigma}_{\mathbf{k}'}^-+g\sqrt{N}\sum_{\mathbf{k}'}(\hat{\sigma}_{\mathbf{k}'}^+\hat{a}_{\mathbf{k}'}+\hat{\sigma}_{\mathbf{k}'}^-\hat{a}_{\mathbf{k}'}^\dagger).
   \label{eq:HGTCk}
\end{equation}
Note that unlike the photonic $\mathbf{k}$ vectors, $\mathbf{k}'=(k_x',k_y'), k_x',k_y'=2\pi l/L,l=0,...,\sqrt{N}-1$ is characterized by a finite cutoff $2\pi/d$ associated with the finite spacing between molecules. According to cavity quantum electrodynamics, the single-molecule coupling $g$ scales inversely with the square root of the mode volume \cite{CohenTannoudji1998}. Consequently, the collective splitting $2g\sqrt{N}$ remains approximately constant when the molecular areal density is fixed while the lateral cavity area is varied. When $L$ is large enough, the sums over $\mathbf{k},\mathbf{k}'$ can be converted to integrals, leading to
\begin{align}
    \hat{H}_{GTC}&=\frac{A}{(2\pi)^2}\int_{\mathbf{k}'\in\mathcal{R}_{\mathbf{k}'}}\left[\hbar\omega(\mathbf{k}')\hat{a}(\mathbf{k}')^\dagger\hat{a}(\mathbf{k}')+E_{xg}\hat{\sigma}^+(\mathbf{k}')\hat{\sigma}^-(\mathbf{k}')+g\sqrt{N}\left(\hat{a}^\dagger(\mathbf{k}')\hat{\sigma}^-(\mathbf{k}')+\hat{a}(\mathbf{k}')\hat{\sigma}^+(\mathbf{k}')\right)\right]d^2k'\notag\\
    &+\frac{A}{(2\pi)^2}\int_{\mathbf{k}\notin\mathcal{R}_{\mathbf{k}}}\hbar\omega_{\mathbf{k}}\hat{a}_{\mathbf{k}}^\dagger\hat{a}_{\mathbf{k}}d^2k,
    \label{eq:Hpolarpphoto}
\end{align}
where $\mathcal{R}_{\mathbf{k}'}=\{(k_x',k_y'):-\pi/d\leq k_x',k_y'<\pi/d$\}. The last term in Eq. \eqref{eq:Hpolarpphoto} corresponds to photon modes $(k_x, k_y)$ outside the range $\mathcal{R}_{\mathbf{k}}$ of interest. Because these modes do not couple to the molecular excitations, they will be ignored hereafter. Disregarding this term and truncating the photonic Hilbert space for each planar mode $\mathbf{k}$ to its two lowest states, $|0_{\mathbf{k}}\rangle$ and $|1\rangle_{\mathbf{k}}$, the partition function is found to be (see details in Sec. II in SI)

\begin{figure}[!htp]
\includegraphics[width=0.8\textwidth]{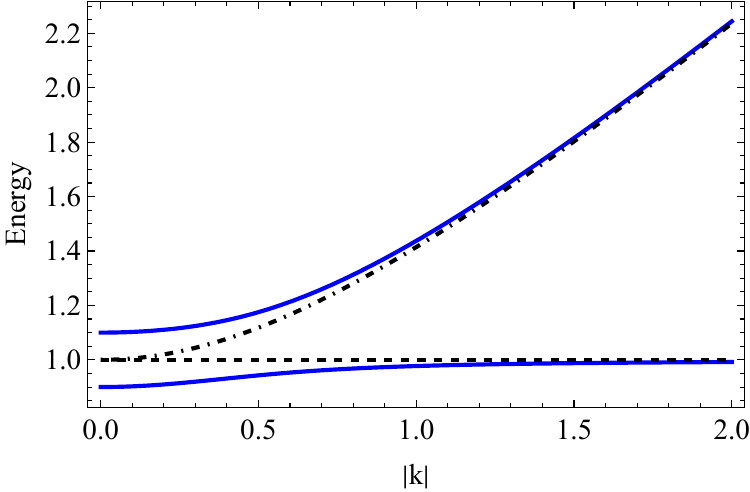}
\caption{Dispersion relation of the upper and lower polaritons [cf. Eq. \eqref{eq:polardisp}]. The Rabi splitting is fixed at $2g\sqrt{N}=0.2$. The dashed black line denotes the molecular electronic transition energy $E_{xg}$, and the dotdashed line denotes the planar photon dispersions.}
\label{fig:1}
\end{figure}

\begin{align}
    \log \mathcal{Z}&=\frac{A}{(2\pi)^2}\int_{-\frac{\pi}{d}}^{\frac{\pi}{d}}\int_{-\frac{\pi}{d}}^{\frac{\pi}{d}}\log\left[1+e^{-\beta E^+(\mathbf{k})}\right]+\log\left[1+e^{-\beta E^-(\mathbf{k})}\right]d^2k,
    \label{eq:Zpolardark}
\end{align}
in which polaritonic dispersions are given by, 
\begin{equation}
    E^\pm(\mathbf{k})=\frac{1}{2}\left[\hbar\omega(\mathbf{k})+E_{xg}\pm\sqrt{(E_{xg}-\hbar\omega(\mathbf{k}))^2+4g^2N}\right].
    \label{eq:polardisp}
\end{equation}
Note that throughout this work, we restrict the Hilbert space to the low-excitation regime, retaining only the ground and first excited states for each molecular and photonic mode. The resulting partition function should therefore be viewed as an effective low-excitation approximation.

\begin{figure}[!htp]
\includegraphics[width=0.8\textwidth]{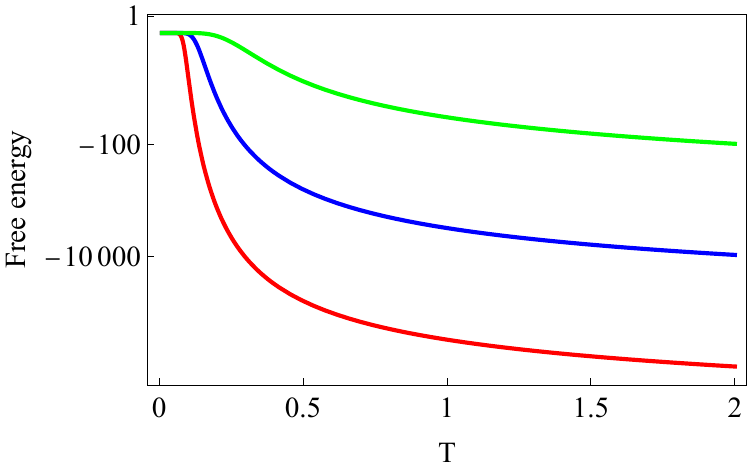}
\caption{Temperature dependence of the free energy for a cavity of fixed lateral size $L=100$. The red, blue and green curves correspond to lattice spacing $d=0.1,1$ and $10$, respectively. The Rabi splitting is fixed at $2g\sqrt{N}=0.2$.}
\label{fig:2}
\end{figure}

The dispersion relation Eq. \eqref{eq:polardisp} of upper $(+)$ and lower $(-)$ polaritons - hybridized states formed by molecular excitations and cavity photons within the single-exciton subspace- is shown in Fig. \ref{fig:1} for the resonant condition $\hbar\omega(\mathbf{k}=0)=E_{xg}$, corresponding to zero detuning between the molecular transition and the normally incident cavity mode. Figure S1 in the SI shows the corresponding dispersion for a larger Rabi splitting, which leads to a larger gap at $|k|=0$. As seen from the dispersion, in the long-wavelength limit, $|k|\ll1$, the photonic and molecular components are maximally hybridized. At larger in-plane wavevectors, the photon and molecular energies separate, and the upper and lower polaritons approach pure photon-like and molecule-like excitations, respectively.

Figure \ref{fig:2} shows the temperature dependence of the free energy $F=-k_BT\log \mathcal{Z}$. The conversion $\sum_\mathbf{k}\rightarrow A/(2\pi)^2\iint d\mathbf{k}$ is valid in the large $L$ limit. At fixed cavity size, denser molecular clusters yield a larger free-energy magnitude at a given temperature.

\section{Molecular chemical reactivity}
We next quantify a simple thermally averaged measure of chemical activity that depends only on the molecular electronic subspace. We adopt a minimal model in which molecules are chemically active whenever they are electronically excited,
\begin{align}
R(E)=
\begin{cases}
R_0, & E>E_G,\\
0, & \text{otherwise},
\end{cases}
\end{align}
where $E_G$ denotes the ground-state energy. We emphasize that this definition is a population-based proxy: at the level of the full molecular ensemble, a reaction can occur whenever the system occupies any state with nonzero molecular excitation character. Accordingly, we define the molecular chemical reactivity $\langle R\rangle$ for a general molecular system
\begin{equation}
    \langle R\rangle=\frac{\int dE\rho(E)e^{-\beta E}R(E)}{\int dE\rho(E)e^{-\beta E}}.
    \label{eq:chemreac}
\end{equation}
where $\beta=1/(k_BT)$ and $\rho(E)$ is the density of energy states \footnote{Obviously, this is a very simplified reaction model that assumes that the reaction rate does not depend on the energy in the excited molecular manifold.}. 
Note that the definition of thermal-averaged reactivity is based on system-level states: A reaction can occur as long as the system is not in the ground state. Here, it is important to note that $\langle R\rangle$ is not a microscopic reaction rate constant, but rather a thermally averaged proxy for excitation-enabled chemical activity.  For a collection of identical molecules, which, upon being regarded as $N$ two-level systems, the corresponding reactivity is
\begin{align}
    \langle R\rangle_M=R_0\frac{(1+e^{-\beta E_{xg}})^N-1}{\left(1+e^{-\beta E_{xg}}\right)^N}.
    \label{eq:outR}
\end{align}
This thermally averaged reactivity depends only on the total number of molecules, regardless of the dimension and geometry of the molecular cluster. Its dependence on $N$ is shown in Fig. S2 in the SI: larger systems exhibit greater chemical activity at higher temperatures.

On the other hand, the molecular reactivity of a planar molecular layer in the microcavity is
\begin{equation}
    \langle R\rangle=R_0\frac{\prod_{\mathbf{k}}\left[1+|c_M^+(\mathbf{k})|^2e^{-\beta E^+(\mathbf{k})}\right]\left[1+|c_M^-(\mathbf{k})|^2e^{-\beta E^-(\mathbf{k})}\right]-1}{\mathcal{Z}},
    \label{eq:GTCCR}
\end{equation}
where $\mathcal{Z}$ is the partition function \eqref{eq:Zpolardark} for the canonical ensemble. 
We emphasize that since polaritons are hybridized light-matter states, while the chemical reactivity only adheres to molecular states, the Hopfield coefficients reflecting molecular component in polaritons,
\begin{align}
    |c_M^-(\mathbf{k})|&=\left[1+\left(\frac{E^-(\mathbf{k})-E_{xg}}{g\sqrt{N}}\right)^2\right]^{-1/2},\\
    |c_M^+(\mathbf{k})|&=\left[1+\left(\frac{g\sqrt{N}}{E^+(\mathbf{k})-\hbar\omega(\mathbf{k})}\right)^2\right]^{-1/2},
    \label{eq:eigenweight}
\end{align}
should be taken into account and associated with polaritonic states (see full derivations in Sec. II in SI). Consistent with Fig. \ref{fig:1}, the upper polariton becomes photon-like and the lower polariton becomes molecule-like at large wavevector. Accordingly, $|c_{M}^+(\mathbf{k})|\rightarrow0$ whereas $|c_{M}^-(\mathbf{k})|\rightarrow1$ as $|k|\rightarrow\infty$. Moreover, if $g\rightarrow0$, corresponding to vanishing light-matter coupling, the same limiting behavior is recovered.

\begin{figure}[!htp]
\includegraphics[width=0.8\textwidth]{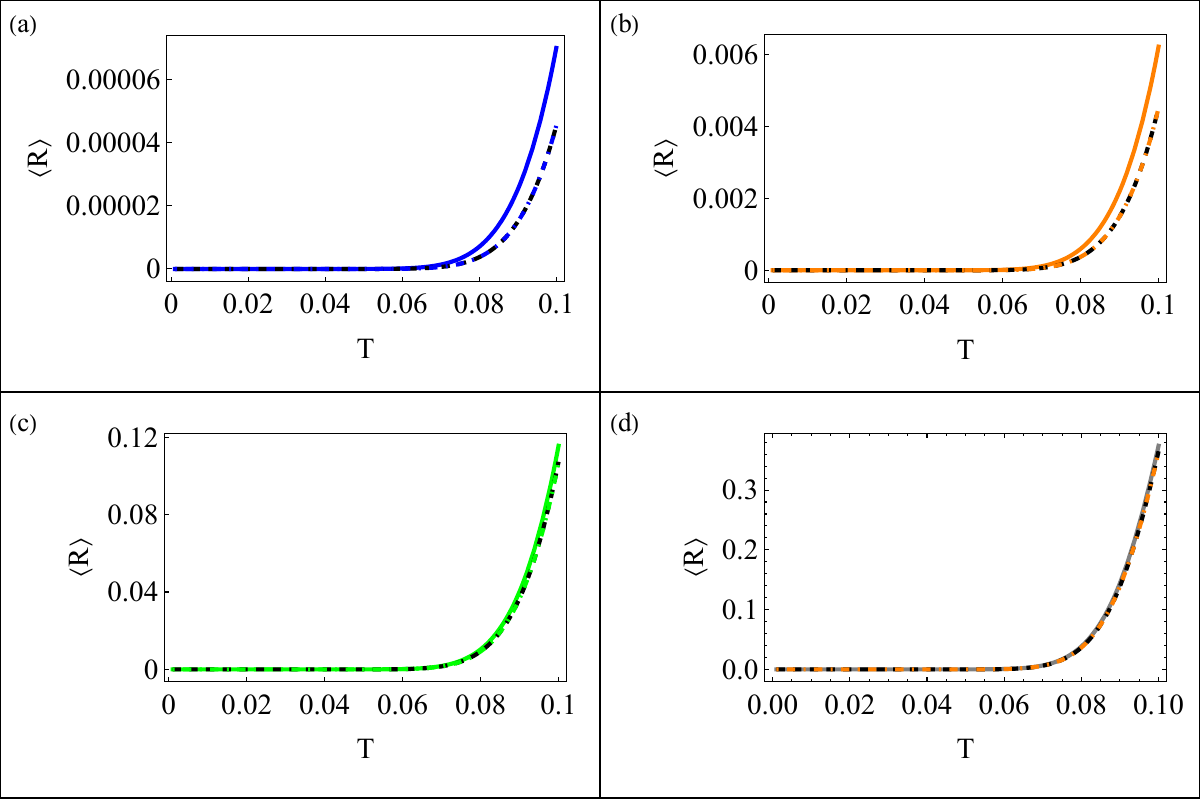}
\caption{Molecular chemical reactivity, in units of $R_0$, for molecules inside the cavity [Eq. \eqref{eq:GTCCR}, solid lines], outside the cavity [Eq. \eqref{eq:outR} dashed lines], and calculations from Eq. \eqref{eq:GTCCR} with $g=0$ (dotdashed lines). Panels (a)-(d) correspond to $N=1,100,2500$, and $10000$, respectively, with fixed cavity size $L=100$. The Rabi splitting is fixed at $2g\sqrt{N}=0.2$. In all panels, the dashed and dotdashed lines overlap.}
\label{fig:3}
\end{figure}

Figure~\ref{fig:3} compares $\langle R\rangle$ for ensembles inside and outside the cavity as a function of temperature and ensemble size. For the parameters considered, the cavity-induced change in $\langle R\rangle$ is small and becomes negligible as $N$ increases. This behavior reflects the increasing contribution of large-$|\mathbf{k}|$ modes, for which the lower polariton becomes predominantly molecular in character (Fig.~\ref{fig:1}). In this limit, the molecular component of the relevant eigenstates - and therefore the population-based chemical activity - approaches that of uncoupled molecules.

We note that Eq. \eqref{eq:GTCCR} is formulated within the single-exciton subspace of the coupled molecule-cavity system. In the limit $g\rightarrow0$, the partition function $\mathcal{Z}$ factorizes into the product of the partition functions for the bare molecular ensemble and the uncoupled cavity photons. However, the numerator in Eq. \eqref{eq:GTCCR} retains only the molecular contribution of the hybridized states and does not include states with higher photonic excitations. As a result, the chemical reactivity obtained from Eq. \eqref{eq:GTCCR} may slightly underestimate the true $g\rightarrow0$ molecular limit. In the temperature range shown here, however, the difference is numerically negligible. In other words, this discrepancy is not apparent because higher excited states are negligibly populated.

In typical Fabry--P\'{e}rot cavities, very large molecular ensembles are coupled to each cavity mode ($N \sim 10^8$--$10^9$) \cite{Dunkelberger2022}, suggesting that any modification of this population-based activity will be modest. In contrast, plasmonic nanocavities can access strong coupling with substantially smaller ensembles, where changes in mode structure and collective coupling can have a more noticeable effect, although the absolute thermal excitation probability remains small under ambient conditions.

\begin{figure}[!htp]
\includegraphics[width=0.8\textwidth]{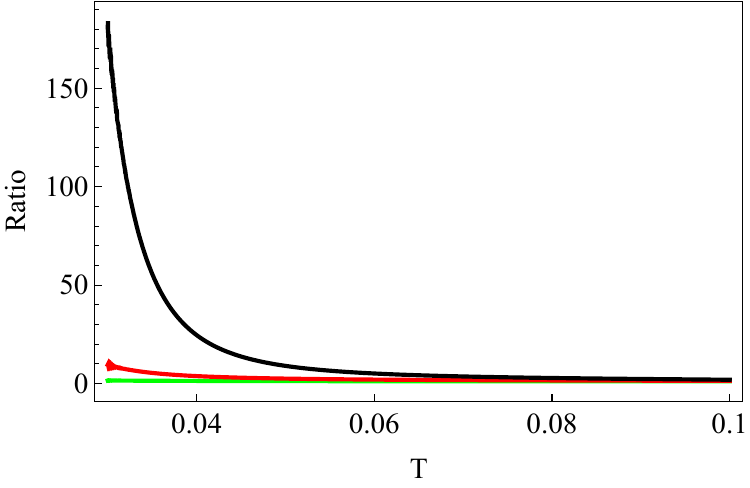}
\caption{Ratio of the molecular chemical reactivity for $N=2500$ molecules inside a cavity of fixed cavity size $L=100$ relative to that outside the cavity. The reactivity inside the cavity is calculated from  Eq. \eqref{eq:GTCCR}. Green, red and black curves correspond to $2g\sqrt{N}=0.2,0.4$ and $0.6$ respectively.}
\label{fig:4}
\end{figure}

Figure \ref{fig:4} compares the ratio of the chemical reactivity of molecules hybridized with cavity photons to that outside the cavity for different coupling strengths (see Fig. S4 in SI for additional examples at other values of $N$). Molecules strongly coupled to the cavity field are more chemically active than bare molecules in this model. Specifically, for a cavity of fixed size containing a fixed number of molecules (i.e. fixed molecular density within a layer), increasing the light-matter coupling strength, and thus the Rabi splitting, enhances overall reactivity, particularly at low temperatures. These results apply within the strong-coupling regime. For sufficiently large Rabi splitting, the system may enter the ultrastrong or deep-strong coupling regime \cite{Lamata2019}.

Overall, our calculations indicate that the thermal population effects underlie the cavity-induced modification of the chemical activity in this model. This mechanism differs from explanations based on barrier modification along a reaction coordinate or dynamical friction/caging effects \cite{Litinskaya2004}, which typically predict very broad spectral features without a clear enhancement of reactivity.

\section{Conclusion}
We have analyzed the thermodynamics of Frenkel-type excitations collectively coupled to planar-cavity photon modes within a generalized Tavis--Cummings description that retains the in-plane dispersion. Using a population-based, excitation-enabled proxy for chemical activity, we find that cavity-induced modifications are most apparent for small ensembles (low areal density within a given mode volume) and increase with collective coupling strength (Rabi splitting). For large ensembles, the activity approaches the uncoupled-molecule result because the dominant contribution arises from large-$|\mathbf{k}|$ states with predominantly molecular character.

Our analysis employs a truncated Hilbert space containing the ground state and single-excitation sectors for each mode, which is appropriate in the low-excitation regime relevant to the present thermodynamic estimates. Extensions that incorporate energetic disorder and more realistic, coordinate-dependent reaction kinetics will be needed to assess cavity effects on specific chemical barriers and rate constants. We also note that excitons and photons are bosons, so hybrid Frenkel exciton-polariton systems might experience condensation under appropriate conditions \cite{Popov1988,Keeling2020}. A full treatment of condensation physics will be addressed in future studies.

\section*{Supplementary Material}
See the supplementary material for the Fourier transform of the generalized Tavis-Cummings model, statistical mechanics of the cavity-molecule systems discussed in the paper, and more figures complementary to the results presented in Sec. III.

\section*{Acknowledgements}
B.C. acknowledges the financial support of the National Natural Science Foundation of China (No. 12404232), start-up funding from the Chinese University of Hong Kong, Shenzhen (No. UDF01003468) and the Shenzhen City “Pengcheng Peacock” Talent Program. AN's research is supported by the European Research Council under ERC-2024-SyG101167294; UnMySt.

\section*{Statements and Declarations}

\subsection*{Conflict of interest}
The author declares no conflict of interest.

\section*{Data availability}
The data that support the findings of this study are available from the corresponding author upon reasonable request.

\bibliography{reference}

\end{document}